# Vector mixed-gap surface solitons

Yaroslav V. Kartashov, Fangwei Ye, and Lluis Torner

*ICFO-Institut de Ciencies Fotoniques, and Universitat Politecnica de Catalunya, Mediterranean Technology Park, 08860 Castelldefels (Barcelona), Spain*
*Yaroslav.Kartashov@icfo.es*

**Abstract:** We elucidate the properties of mixed-gap vector surface solitons supported by the interface between a uniform medium and an optical lattice imprinted in a Kerr-type nonlinear media. The components of such mixed-gap solitons emerge from different gaps of lattice spectrum and their mutual trapping results in the formation of stable vector states. The unstable soliton component is stabilized by the cross-coupling with the stable component. We show that vector mixed-gap surface solitons exhibit a new combination of properties of vectorial surface waves and gap solitons.

Cross-coupling between several light waves can considerably enrich the dynamics of their propagation. Such coupling provides a strong stabilizing mechanism, which results in the formation of light patterns that often do not even exist in scalar settings. Vector solitons forming due to the coupling of two waves have been encountered in a number of geometries, including uniform media with both coherent [1-4] and incoherent [5-8] interactions. On the other hand, several previous investigations have shown that a transverse periodic modulation of the refractive index strongly affects the properties of vector solitons. For example, two-dimensional optical lattices [9] and arrays of weakly coupled waveguides support strongly localized vectorial modes that have no analogs in uniform media [10-12]. Similarly, mixed-gap vector solitons whose components emerge from different gaps of the Floquet-Bloch spectrum of the periodic structure have been introduced recently [13,14]. Optical lattices support a variety of other stable vector solitons [15].

The presence of interfaces between different materials substantially modifies the properties of both scalar and vector solitons. It was shown recently that the interface between a waveguide array and a uniform material allows formation of new types of scalar surface solitons [16,17]. Prior to [17], optical surface waves were also observed at the nonlinear interface of a photorefractive crystal with diffusion nonlinearity [18], as well as at linear interfaces of periodic media [19]. Observation of highly nonlinear surface waves at interfaces of natural materials is a challenge due to the large powers required for their excitation (see [20-23] for reviews). Interfaces of periodic media support also gap scalar solitons that were predicted theoretically in [24] and observed experimentally in [25]. Gap solitons may also exist at surfaces that are periodically modulated in the direction of light propagation [26].

Besides scalar solitons, interfaces of periodic structures support vector surface waves. The simplest vector surface solitons were studied in [27], for the case of coherently interacting components that emerge from a semi-infinite gap. Nevertheless, the most interesting situation is encountered when the components of the vector surface soliton originate from different gaps. Several examples of such solitons were obtained recently at the interfaces of discrete binary waveguide arrays with focusing nonlinearity [28]. In this paper we study mixed-gap vector surface waves arising at the interface of a uniform medium and an optical lattice imprinted in a Kerr nonlinear medium, for focusing and defocusing nonlinearities. We reveal that cross-coupling results in the formation of stable vector states, where instability of one of the components is suppressed by the interaction with the stable component.

We consider the propagation of two mutually incoherent laser beams at the interface between a cubic uniform medium and a periodic optical lattice. For concreteness, here we assume that the strength of the cross-phase-modulation is equal to that of the self-phase-modulation. In this case the light propagation is described by two coupled nonlinear Schrödinger equations for the dimensionless complex amplitudes $q_{1,2}$:

$$i\frac{\partial q_{1,2}}{\partial \xi} = -\frac{1}{2}\frac{\partial^2 q_{1,2}}{\partial \eta^2} + \sigma q_{1,2}(|q_1|^2 + |q_2|^2) - pR(\eta)q_{1,2}. \qquad (1)$$

Here the transverse $\eta$ and longitudinal $\xi$ coordinates are scaled in terms of the beam width and diffraction length, respectively; the parameter $\sigma = \mp 1$ stands for focusing/defocusing; $p$ is the lattice depth, and the function $R(\eta) = 1 - \cos(\Omega\eta)$ for $\eta \geqslant 0$ and $R(\eta) \equiv 0$ for $\eta < 0$ describes the transverse refractive index profile. The quantity $\Omega$ is the lattice frequency. Eqs. (1) conserve the total energy flow

$$U = U_1 + U_2 = \int_{-\infty}^{\infty}(|q_1|^2 + |q_2|^2)d\eta. \qquad (2)$$

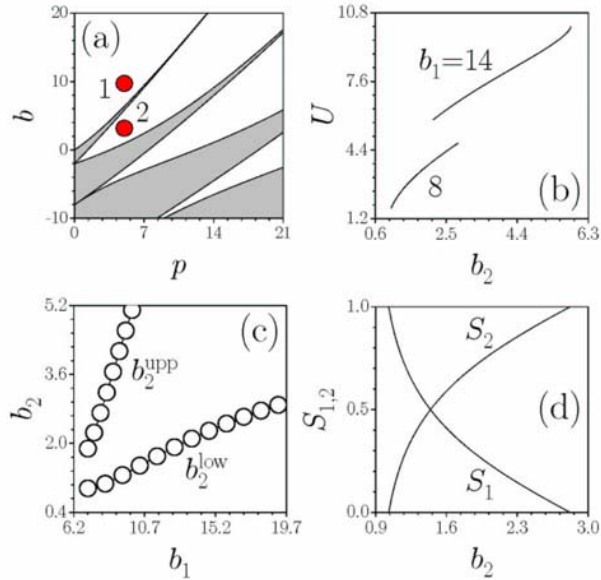

Fig. 1. (a) Schematic representation of the loci of the nonlinear propagation constants of the vector surface soliton components in the band-gap lattice spectrum. (b) Total energy flow vs $b_2$ at $p = 5$. (c) Domain of existence of vector surface solitons at the $(b_1, b_2)$ plane at $p = 5$. (d) Energy sharing between soliton components vs $b_2$ at $b_1 = 8$ and $p = 5$. Focusing medium. Here and throughout the paper we set $\Omega = 4$.

Interfaces between uniform and periodic media can be fabricated technologically, e.g. by etching array of waveguides on a nonlinear substrate at $\eta > 0$ [11,12]. Optical induction offers a potentially powerful alternative. In this case, a periodic pattern is imprinted in a photo-refractive crystal by using several interfering plane waves and subsequently erased at $\eta < 0$ with an intense green wave propagating in the direction orthogonal to $\eta$ and $\xi$ axes.

We search for vector soliton solutions of Eq. (1) in the form $q_{1,2} = w_{1,2}(\eta)\exp(ib_{1,2}\xi)$, where $w_{1,2}$ are real functions and $b_{1,2}$ are propagation constants. We are interested in solitons whose maxima seat on the lattice channel nearest to the interface. The periodicity of the lattice imposes restrictions on the available values of propagation constants of localized solitons, that should fall into gaps of the Floquet-Bloch lattice spectrum (white regions in Fig. 1(a)). In the gray regions (bands), lattices can support only delocalized Bloch waves. Surface solitons residing at the interface with the lattice penetrate into the uniform medium too, which imposes the conditions $b_{1,2} \geq 0$ on the propagation constants. When $b_1 = b_2$, the components $w_{1,2}$ have similar shapes and their properties are analogous to those for scalar gap solitons [24]. We assume that $b_1$ and $b_2$ fall into neighboring gaps of the Floquet-Bloch spectrum, so that $w_{1,2}$ feature different structures. In a focusing medium ($\sigma = -1$) the components of the simplest mixed-gap vector soliton emerge from the semi-infinite and the first finite gaps (see Fig. 1(a)). Representative profiles are shown in Figs. 2(a)-2(c).

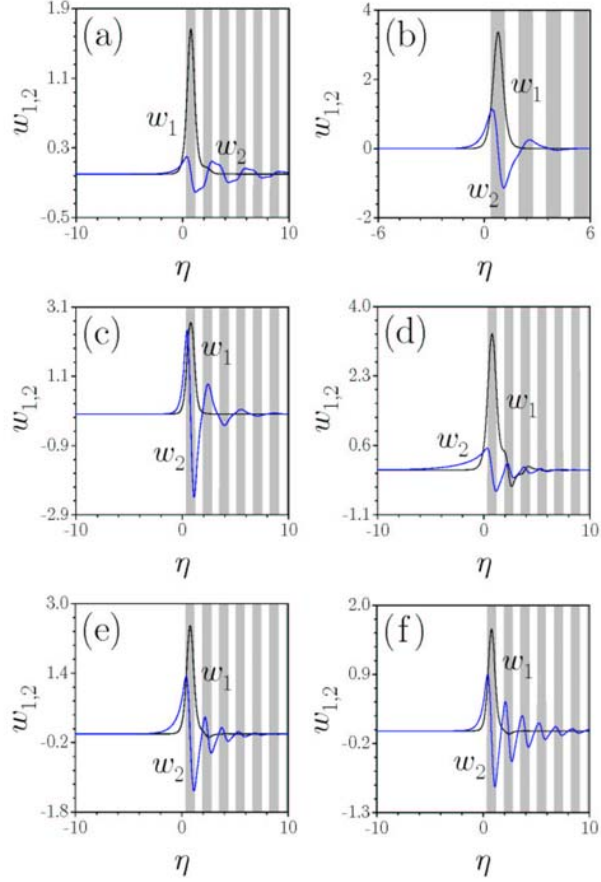

Fig. 2. Profiles of mixed-gap vector surface solitons at different propagation constant values. (a) $b_1 = 8$, $b_2 = 1.05$, (b) $b_1 = 14$, $b_2 = 3$, (c) $b_1 = 14$, $b_2 = 5.7$, (d) $b_1 = 5.8$, $b_2 = 0.1$, (e) $b_1 = 9$, $b_2 = 1$, (f) $b_1 = 12$, $b_2 = 2.6$. Panels (a)-(c) correspond to $p = 5$ and focusing medium. Panels (d)-(f) correspond to $p = 10$ and defocusing medium. In gray regions $R(\eta) > 1$, while in white regions $R(\eta) \leq 1$.

Physically, each component of the vector gap surface soliton forms by the nonlinear coupling between the incident and the reflected waves having equal wavevectors along the longitudinal coordinate (i.e. $\xi$) and opposite wavevectors in the transverse coordinate (i.e. $\eta$), when both of them experience Bragg scattering from the periodic structure. The symmetry of the soliton components reflects the symmetry of the Bloch waves near the corresponding gap edges. Such symmetry is determined by the effective sign of the diffraction required for the formation of localized states in focusing (normal diffraction) and defocusing (anomalous diffraction) media. Cross-modulation coupling between such components results in their local distortion in the regions where the amplitudes are highest and weakly affect the soliton tails, especially when the degree of localization of the two components differs significantly.

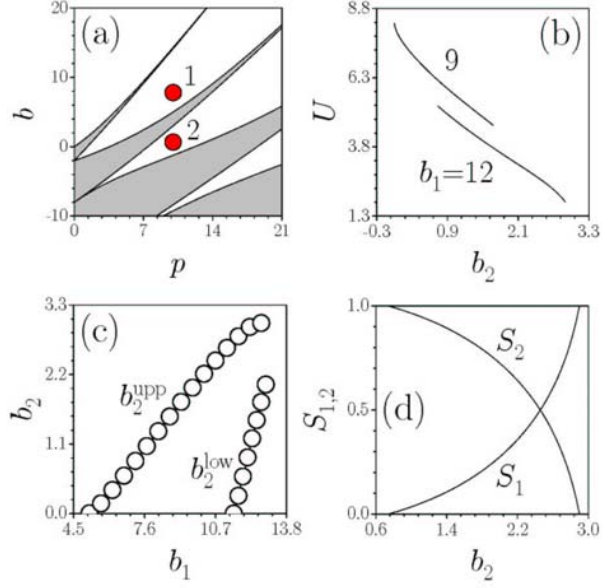

Fig. 3. (a) Representation of the loci of the propagation constants of the vector surface soliton components in the band-gap lattice spectrum. (b) Total energy flow vs $b_2$ at $p=10$. (c) Domain of existence of vector surface solitons at the $(b_1,b_2)$ plane at $p=10$. (d) Energy sharing between soliton components vs $b_2$ at $b_1=12$ and $p=10$. Defocusing medium.

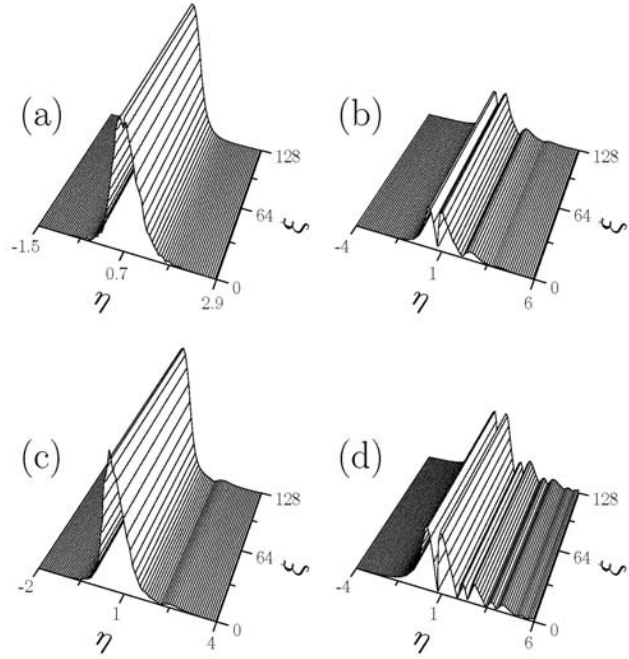

Fig. 4. Stable propagation of mixed-gap surface vector solitons in the presence of white input random perturbations with variance $\sigma_{\text{noise}}^2=0.01$ added to the exact soliton profiles. Panels (a) and (b) show the modulus of the first and second components for a soliton in focusing media with $b_1=14$, $b_2=3$, $p=5$. Panels (c) and (d) show the modulus of the first and second components for a soliton in defocusing media with $b_1=9$, $b_2=1$, $p=10$. Vertical scales are the same for panels (a) and (b) and for panels (c) and (d).

In the case $\sigma = -1$ the component $w_1$ is always positive, while $w_2$ exhibits multiple oscillations at $\eta > 0$. Both components decay exponentially as $\eta \to -\infty$. For fixed $p$ and $b_1$, the energy flow $U$ is a monotonically increasing function of $b_2$ (Fig. 1(b)). There exist a lower $b_2^{\text{low}}$ cutoff and an upper $b_2^{\text{upp}}$ cutoff on $b_2$ for the existence of composite vector solitons. Figure 1(c) shows the domain of existence on the $(b_1, b_2)$ plane for $p = 5$. We found vector solitons for arbitrary large values of $b_1$. At large $b_1$ the component $w_1$ is narrow and its amplitude is much higher than that of $w_2$. The upper cutoff $b_2^{\text{upp}}$ coincides with the upper edge of the first gap ($b_2^{\text{upp}} = 5.834$ at $p = 5$) and for $b_2 \to b_2^{\text{upp}}$ the second component deeply penetrates into the lattice region thus approaching a delocalized Bloch wave. In the upper cutoff $w_1$ remains well localized (Fig. 2(c)). As $b_2 \to b_2^{\text{low}}$, the second component vanishes, while its transverse extent depends on position of $b_2^{\text{low}}$ in the first gap. For $b_1 \leqslant 10.6$, the upper cutoff $b_2^{\text{upp}}$ abruptly decreases when $b_1$ also decreases. In this regime, $w_1$ vanishes in the upper cutoff. Figure 1(d) shows the energy sharing $S_{1,2} = U_{1,2}/U$ between the vector soliton components for this parameter range. With further decrease of $b_1$ to 7.1, the lower cutoff $b_2^{\text{low}}$ gradually approaches the lower edge of first gap, so that $w_2$ at $\eta > 0$ and $b_2 \to b_2^{\text{low}}$ the soliton closely resembles a Bloch wave corresponding to this gap edge (Fig. 2(a)).

Notice that for $b_1 \leqslant 7.1$ we did not find solitons for which $w_2$ has a maximum in the first lattice channel. For $b_1 \to 7.1$ strong secondary peaks appear in $w_1$, while the maximum of $w_2$ shifts into the lattice so that the two components decouple. Similar domains of existence were encountered for various lattice depths. For small enough $p$, when the propagation constant corresponding to the lower edge of the first gap becomes negative, the domain of existence shown in Fig. 1(b) is bounded from below by the line $b_2^{\text{low}} = 0$.

Defocusing media ($\sigma = 1$) also support mixed-gap solitons. In this case, vector soliton components emerge from the first and the second finite gaps of the Floquet-Bloch spectrum (Fig. 3(a)). At fixed $b_1$ and $p$ the energy flow is a monotonically decreasing function of $b_2$ (Fig. 3(b)). At $p = 10$ the domain of existence of composite states appears at $b_1 = 5.008$ corresponding to the lower edge of the first gap (Fig. 3(c)). In the vicinity of this point, $w_1$ turns out to be weakly localized and closely resembles a Bloch wave in the region $\eta > 0$. The lower cutoff is given by $b_2^{\text{low}} = 0$, and as $b_2 \to b_2^{\text{low}}$, $w_2$ deeply penetrates into the uniform medium (Fig. 2(d)) being well localized inside the lattice. In contrast to the case of focusing media, in the upper cutoff $w_2$ vanishes. With increase of $b_1$ the energy flow of the first component at $b_2 \to b_2^{\text{low}}$ gradually decreases. For $b_1 \geqslant 11.52$ the lower cutoff $b_2^{\text{low}}$ abruptly increases with $b_1$, while the well-localized $w_1$ component completely vanishes at $b_2 \to b_2^{\text{low}}$. The upper cutoff $b_2^{\text{upp}}$ increases with $b_1$ and at $b_1 = 12.5$ it reaches the upper edge of the second gap given by $b_2 = 3.015$. The closer $b_2^{\text{upp}}$ to this value the stronger the expansion of the small-amplitude $w_2$ component into the lattice region near upper cutoff (compare $w_2$ in Figs. 2(f) and 2(e)). At $b_1 > 12.5$ the $w_2$ component does not vanish at $b_2^{\text{upp}}$ and $w_2$ transforms into a delocalized wave in the lattice region. Finally, the composite vector states cease to exist for $b_1 \approx 12.9$, since the maximum of $w_2$ shifts into the lattice region and $w_{1,2}$ decouple. The energy sharing between components of the vector soliton vs $b_2$ is shown in Fig. 3(d). One concludes that the rich internal structure of the vector surface waves is determined by the location of the propagation constants $b_{1,2}$ in the band-gap spectrum, locations which are affected by the presence of the interface.

To analyze the dynamical stability of the mixed-gap vector surface solitons we solved Eq. (1) numerically with input conditions $q_{1,2}|_{\xi=0} = w_{1,2}(1 + \rho_{1,2})$, where $\rho_{1,2}(\eta)$ describes a Gaussian random perturbation with variance $\sigma_{\text{noise}}^2$. We verified that in the absence of $w_1$, the scalar surface solitons that emerge from the first gap in focusing media and from the second gap in defocusing media (i.e. those having internal structure similar to that of $w_2$ in Fig. 2) are strongly unstable. In contrast, scalar solitons having internal structure of $w_1$ can be stable in large parts of their existence domain. Intuitively, this suggests the possibility to stabilize composite vector solitons made of a relatively weak $w_2$ and a strong $w_1$. This occurs in wide regions near the lower cutoff $b_2^{\mathrm{low}}$ in focusing media and near the upper cutoff $b_2^{\mathrm{upp}}$ in defocussing media. Thus, at $\sigma = -1$ and $p = 5$ vector solitons are stable for $b_2 \in [1.029, 1.43]$ at $b_1 = 8$, while at $b_1 = 14$ the stability domain is given by $b_2 \in [2.16, 5.29]$. We found that the stability domain substantially extends with $b_1$ and the upper edge of the instability domain gradually approaches $b_2^{\mathrm{upp}}$.

Stable propagation of the vector surface soliton depicted in Fig. 2(b) is shown in Figs. 4(a) and 4(b). At $\sigma = 1$ the stability domain at $p = 10$ is given by $b_2 \in [0.58, 1.675]$ for $b_1 = 9$ and $b_2 \in [2.45, 2.9]$ for $b_1 = 12$. The width of the stability domain is maximal for $b_1 \sim 9$. Stable propagation of the vector soliton depicted in Fig. 2(e) is shown in Figs. 4(c) and 4(d). In all cases, the stable vector solitons retain their input structure over indefinitely long distances even in the presence of considerable input random perturbations. However, in both cases, increasing the weight of the $w_2$ component results in the destabilization of the vector soliton. All instabilities encountered for vector surface solitons are oscillatory. They result in progressively increasing oscillations of amplitudes of the components of unstable soliton, emission of radiation into the lattice and decay into stable solitons. Here we present results of stability analysis for only one lattice depth, but we verified by a detailed analysis that different lattice depths yield qualitatively similar results.

In summary, we reported a detailed analysis of the properties and stability of new vector solitons supported by the interface of a uniform medium and optical lattice. The components of such vector solitons emerge from different gaps of the Floquet-Bloch lattice spectrum, and exhibit different internal structures and stability properties. Nevertheless, we revealed that their cross-coupling leads to formation of new types of completely stable, mixed-gap surface vector solitons.

**Acknowledgements**

This work has been supported in part by the Government of Spain through the Ramon-y-Cajal program and through the grant TEC2005-07815/MIC.